\documentstyle[12pt]{article}
\newcommand{\beq}{\begin{equation}}
\newcommand{\eeq}{\end{equation}}
\newcommand{\bra}{\begin{array}}
\newcommand{\era}{\end{array}}

\title{ The Moyal Bracket in the Coherent States framework }
\author{ M. Daoud$^{1 *}$ and E. H. El Kinani$^{2 *}$}
\date{}
\begin{document}
\maketitle
\begin {center} {The Abdus Salam International Centre for Theoretical Physics,
 ICTP-Strada costiera 11, 34100 Tieste  Italy.}
\end {center}

\vskip 1cm

{\bf {Abstract}}.\\

The star product and Moyal bracket  are introduced using the
coherent states corresponding to quantum systems with non-linear
spectra. Two kinds of coherent state are considered. The first
kind is the set of Gazeau-Klauder coherent states and the second
kind are constructed following the Perelomov-Klauder approach.
The particular case of the harmonic oscillator is also discussed.  \\

\vskip 8cm

\hrule

$^{(2)}$E-mail: hkinani@ictp.trieste.it or el-kinani@fste.ac.ma \\
$^{(*)}$Permanent address:\\
$^{(1)}$ LPMC,  D\'epartement de Physique, Facult\'e des
Sciences B.P 28/S, Universit\'e Ibn Zohr Agadir, Morocco \\
$^{(2)}$ GMP,  D\'epartement de Math\'ematiques, Facult\'e des
Sciences et Technique, Boutalamine B.P 509,  Errachidia, Morocco. \\

\newpage
\section{Introduction}

\hspace{.3in} In classical mechanics, observables are smooth
functions on phase space, which constitute a Poisson algebra,
while in quantum mechanics, the observables constitute a
non-commutative associative algebra. Deformation quantization is
the basis of one of the important attempts aiming to construct a
quantum system starting from a classical mechanics system.  It is
required that the quantum system obtained must go over into the
original classical one in the limit $\hbar \to 0$ where $\hbar$ is
Plank's constant. In recent times, a deformation quantification
has bee explored in several context: in  string theory approach to
noncommutative geometry $\cite{1}$, Matrix Models $\cite{2}$, the
noncommutative Yang-Mills
theories $\cite{3}$ and non-commutative gauge theories $\cite{4}$.\\

Recently, the star product associated with an arbitrary
two-dimensional Poisson structure, using the coherent states on
the complex plane, was introduced $\cite{5}$. It was shown that,
from the coherent states adapted to harmonic oscillator, one
recover easily the well-known Moyal star-product $\cite{6}$. The
deformed coherent states ( \`a la Man'ko et al) $\cite{7}$ were
also considered to provide an associative star-product. Then, it
is clear now that the coherent state formulation give an useful
scheme
to define the star-product in a consistent way.\\

The approach taken in this work is along the lines of
Berezin quantization $\cite{8}$ and relies on coherent states
(Gazeau-Klauder(GK) $\cite{9}$ and Perelomov-Klauder,(PK)
$\cite{10,11}$ ) adapted to an exact solvable systems with
a nonlinear spectrum $\cite{12,13}$. The use of the coherent
states is due to their useful property of overcompletness.
For our purpose, we will consider the coherent states \`a la
Gazeau-Klauder and ones defined following
Perelomov-Klauder approach. Theses constructions leads
as we will see to inequivalent states except for the harmonic
oscillator case.\\

We start by introducing the creation and annihilation operators corresponding
to a quantum systems with non-linear
spectrums of type $e_n = an^2 + bn \:\:\:\ ( n \in {\bf{N}}, a \ge 0, b>0 $).
 For some particular values of $a$ and $b$, one find again well-known
 quantum mechanical
 systems like P$\ddot{o}$schl-Teller potential( see $\cite{12}$),
$x^4$-anharmonic oscillator $\cite{14}$, and standard harmonic
oscillator. Section 3 is devoted to the construction of
Gazeau-Klauder and Perelomov-Klauder coherent states for the above
non-linear quantum systems. Their properties ( resolution to unity
and analytical representations) are also presented. In  section 4,
the Gazeau-Klauder coherent states ( eigenstates  of the
annihilation operator) leads easily to the definition of
star-product and Moyal bracket on the complex plane. However, when
one deals with PK coherent states, the previous definition becomes
nontrivial due to the fact that this kind of coherent states are
not eigenstates of the annihilation operator. To overcome this
difficulty, we introduce a new annihilation operator that
diagonalizes the PK states. Concluding remarks are given in the
last section.

\section{Non-linear quantum  spectrums}
\hspace{.3in}Choose  a Hamiltonian $H$ with a discrete spectrum which is bounded below, and
has been adjusted so that $H \ge 0$ . We assume  that the eigenvalues of $H$ are
non-degenerate. The eigenstates $|\psi_n>$  of $H$ are orthonormal vectors and they
satisfy
\beq
H|\psi_n>=e_n|\psi_n>
\eeq
We suppose that $e_n \ge 0$ and verifies $e_{n+1}>e_n$. The energy $e_0$ of the
 ground state $|\psi_0>$  is chosen to be zero. It is well-known that for such system
 one can factorize the Hamiltonian $H$ in terms of creation $a^+$ and annihilation
 $a^-$ operators as follows
\beq
H= a^+ a^-
\eeq
Theses operators acts on the Hilbert space ${\mathcal{H}} = \{|\psi_{n}>, n \in {\bf N} \}$, as

\beq
a^+|\psi_n>=\sqrt{e_{n+1}} |\psi_{n+1}> \:\:\  \mbox{and} \:\:\
 a^-|\psi_n>=\sqrt{e_n} |\psi_{n-1}>,
\eeq
implemented by $a^-|\psi_0>=0$. We define the operator $G$

\beq
 [ a^- , a^+] =G \sim G(N)
\eeq
as the commutator  between $a^-$ and $a^+$ . It is clear, from equation (3),
that the action of $G$ on the state $|\psi_n>$ is given by
\beq
G|\psi_n> = [ a^- , a^+]|\psi_n>=(e_{n+1}-e_n)|\psi_n>
\eeq
The operators $N$ is defined such that
\beq
N|\psi_n> = n|\psi_n>
\eeq
Note that in general the operator $N$ is different from $H$.
They coincides only in the harmonic oscillator case.
 Furthermore, one can verify also  the following commutation relations

\beq
 [ N, a^{\pm}] = \pm a^{\pm}
\eeq
In this letter, as we mentioned before,  we focus our attention of a quantum systems with energy spectrum of
 type
\beq
e_n = a n^2 +b n   \:\:\:\:\ , n=0,1,2,... \:\:\:\:\ a\ge0 \:\ , \:\  b>0
\eeq
This choice covers many interesting situations. Indeed, for $(a=1, b= k+ k')$
, we have  the spectrum of a quantum system evolving in the
P$\ddot{o}$schl-Teller potentials parametrized by $k$ and $k'$  ( $k>1$ and $k'>1$) $\cite{12,13}$
\beq
 H= - \frac{d^2}{dx^2} + V(x)
\eeq
where
\beq
 V(x)= \frac{1}{4} ( \frac{k(k-1)}{ \sin^{2}(x/2)} +\frac{k'(k'-1)}{ \cos^{2}(x/2)})
  \:\:\, \:\:\ 0< x<\pi
\eeq
and $V(x)= \infty$ otherwise (i.e $ x \ge 0  \:\:\ ; \:\:\ x \ge \pi$). This family of potentials is also
called, sometime, the P$\ddot{o}$schl-Teller potentials of the first kind. The latter reduces to other interesting
potentials, which are widely used in solid state and molecular physical, like for instance Scarf and Rosen-Morse
ones ($\cite{12}$ and references quoted therein). The case $(a=1, b=k+k'=2$), correspond to the spectrum of a free
particle trapped in the infinite square-well potential. In the case $(a=\frac{3\epsilon}{2},  b=a+1$), where the parameter
 $\epsilon$ is positive, we have the energy levels of so-called the  $x^4-$anharmonic
 oscillator $\cite{14}$ describing by the Hamiltonian
\beq
 H= a_{0}^+ a_{0}^- + \frac{\epsilon}{4}(a_{0}^- +a_{0}^+)^4 -c_0
\eeq
where $c_0 =\frac{3 \epsilon}{4}-\frac{21 {\epsilon}^2}{2}$ and $a_{0}^+$ , $a_{0}^-$  are annihilation
and creation operator $(\{a_{0}^- , a_{0}^+ \}=1)$ of the harmonic oscillator. This quantum system has been extensively studied since the early 1970
( see review  $\cite{14}$) . Finally, for $a =0$ and $ b=1$, we obtain the standard harmonic oscillator spectrum
which can be also  obtained  from $x^4-$anharmonic system in the limit $ \epsilon \to 0$.

\section {Coherent states}

\hspace{.3in} Coherent states play an important role in many
different context of theoretical and experimental physics,
especially quantum optics $\cite{11}$. This notion was firstly
discovered for the harmonic oscillator and has been extended for
several other potentials in the references $\cite{9,12,13}$ in
which coherent states are defined: (i) as eigenstates of the
annihilation operator, (ii) by acting the displacement operator on
the ground state $|\psi_0>$ and (iii) as states minimizing the
so-called the Robertson-Shrodinger uncertainty relation. The
definitions (i),(ii),(iii) gives different sets of states when one
deal with a quantum system other than the harmonic oscillator. As
we have mentioned above, we investigate the way to construct the
star-product using the coherent states associated with an
arbitrary quantum systems having spectrum of type $e_n=an^2+bn$.
Two types of coherent states will be used. The first set is the
so-called Gazeau-Klauder GK coherent states obtained from the
definition (i). The second type are of Perelomov-Klauder PK
constructed following the definition (ii). Note that the
minimization of the Robertson-Scrodinger uncertainty relation
leads to the so-called generalized intelligent states, which are
not of interest in this work

\subsection{Gazeau-Klauder Coherent states}
\hspace{.3in} Let us denote the  Gazeau-Klauder coherent states  by $|z> , z \in {\bf C}$. They are
defined as eigenstates of the annihilation operator $a^-$
\beq
 a^-|z>=z|z>
\eeq
Decomposing $|z>$ in Hilbert space ${\mathcal{H}}$ basis, and using the action of $a^-$ on the $|\psi_n>$'s given by (3),
we show that the coherent states $|z>$ are as follows
\beq
 |z>= {\mathcal{N}}(|z|^2)^{-1} \sum_{n=0}^{\infty} \frac { (\Gamma (r+1))^{1/2} z^n}{ n!
 (\Gamma (n+r+1))^{1/2}a^{n/2} }|\psi_n>
\eeq
where $r=\frac{b}{a}$ and the normalization constant ${\mathcal{N}}(|z|^2)$ is
\beq
 ({\mathcal{N}}(|z|^2))^{2}
 =_{0}F_{1} (r+1, \frac{|z|^2}{a})
\eeq
The set of states $|z>$ is overcomplete. Indeed, the resolution of
unity
\beq
 \int |z><\bar{z}| d\mu (z,\bar{z}) =I_{\mathcal{H}}
\eeq
is ensured in respect to the measure :

\beq
  d\mu (z,\bar{z}) =\frac{2}{\pi a} I_{r}(\frac{2r}{\sqrt{ a}})K_{r/2}(
  \frac{2r}{\sqrt{ a}}) rdrd\theta    \:\:\:\ , \:\:\:\  z=re^{i\theta}
\eeq The latter formula can be determined in a different ways.
Here, we have used  the approach developed in $\cite{12,13}$. The
kernel (overlapping of two coherent states) is given by \beq
 <z'|z>= \frac{_{0}F_{1} (r+1, \frac{\bar{z'}z}{a})}{(
 _{0}F_{1} (r+1, \frac{|z|^2}{a}) _{0}F_{1} (r+1, \frac{|z'|^2}{a}))^2}
\eeq
The overcompletion of the set \{ $|z>,  z \in \bf{C} $ \} provide a representation
 of any state by the entire function
\beq
  f(z)=( _{0}F_{1} (r+1, \frac{|z|^2}{a}))^{1/2} <\bar{z}|f>
\eeq
In particular, the analytic function corresponding the eigenstates $|\psi_{n}>$ are

\beq
 {\mathcal{F}}_{n}(z)= \frac{ z^{n} \sqrt {\Gamma(r+1)}} {a^{n/2}(n! \Gamma(n+r+1))^{1/2}}
 \eeq
On the set $ \{ {\mathcal{F}}_{n}(z) \}$ the action of the creation and annihilation operators are given by
\beq
a^+= z \:\:\:\ , \:\:\:\ a^-=(z \frac{d^2}{dz^2}+(r+1))\frac{d}{dz},
\eeq
and the operator $G$ acts as
\beq
G=2az \frac{d}{dz}+(a+b)
\eeq
It is easy to see that theses operators act in the functions space  $ \{ {\mathcal{F}}_{n}(z), n\in {\bf N} \} $   as
\beq
 a^+{\mathcal{F}}_{n}(z) =\sqrt{e_{n+1}} {\mathcal{F}}_{n+1}(z) \:\:\,\:\:\
  a^-{\mathcal{F}}_{n}(z) =\sqrt{e_{n}} {\mathcal{F}}_{n-1}(z)  \:\:\,\:\:\
  G{\mathcal{F}}_{n}(z) =(e_{n+1}-e_n) {\mathcal{F}}_{n}(z)
\eeq
This realization will be useful in the sequel of this work when we will introduce  the
star-product approach based on the Gazeau-Klauder Coherent states.

\subsection{Perelomov-Klauder Coherent states}

\hspace{.3in}We recall that the Perelomov-Klauder Coherent are defined by :
\beq
|z>= {\mathcal{D}}(z)|\psi_0>= exp( z a^+ -\bar{z} a^- )|\psi_0>
\eeq
The computation of the action of the displacement operator ${\mathcal{D}}(z)$
on the ground state $ |\psi_0 >$ was done for an arbitrary quantum system  and illustrated for the
P$\ddot{o}$schl-Teller potentials  $\cite{13}$. Note that this result can be also
applied  for a quantum systems possessing energy levels $e_n=an^2 + bn \:\,
(n \in \bf{N}) $ with a minor modifications .Then, one can obtain
\beq
|\zeta>= (1-|\zeta|^2)^{\frac{r+1}{2}} \sum_{n=0}^{\infty} \sqrt{{ \frac{\Gamma(n+r+1)}
 {n!\Gamma(r+1)}}} \zeta^{n} |\psi_n>
\eeq
where $\zeta= \frac{z}{|z|} \tanh (z \sqrt{a})$.
The states $|\zeta>$ satisfies the resolution to unity, namely
\beq
 \int |\zeta><\zeta| d\mu (\zeta, \bar{\zeta}) =I_{\mathcal{H}},
\eeq
in respect to the measure given by

\beq
  d\mu (\zeta, \bar{\zeta}) = \frac{r}{\pi} \frac{d^{2} \zeta}{(1-| \zeta|^{2})^{2}}
\eeq
The kernel $<\zeta'|\zeta>$ is given by

\beq
  <\zeta'|\zeta>  = (1-|\zeta'|^2)^{\frac{r+1}{2}}(1-|\zeta|^2)^{\frac{r+1}{2}}
  \sum_{n=0}^{\infty}{ \frac{\Gamma(n+r+1)}
 {n!\Gamma(r+1)}} (\bar{\zeta}' \zeta)^n
\eeq
The state $|\psi_n>$ is represented analytically by the function

\beq
 {\mathcal{G}}_{n}(\zeta)= \zeta^{n} \sqrt{\frac{\Gamma(n+r+1)} {n! \Gamma(r+1)}}
\eeq
The creation and annihilation operators act in the Hilbert space of analytical
 functions $\{{\mathcal{G}}_{n}(\zeta), n\in {\bf N} \}$ as a first order differential operators

\beq
a^+=  \zeta^{2} \frac{d}{d \zeta}+(r+1)\zeta \:\:\:\ , \:\:\:\
a^-= \frac{d}{d \zeta}
\eeq
 and the operator $G$ acts in the same representation as
\beq
G=2 \zeta \frac{d}{d \zeta}+(r+1)
\eeq
One can verify that
\beq
 a^+{\mathcal{G}}_{n}(\zeta) =\sqrt{e_{n+1}} {\mathcal{G}}_{n+1}(\zeta) \:\:\,\:\:\
 a^-{\mathcal{G}}_{n}(\zeta) =\sqrt{e_{n}} {\mathcal{G}}_{n-1}(\zeta)  \:\:\,\:\:\
  G{\mathcal{G}}_{n}(\zeta) =e_n {\mathcal{G}}_{n}(\zeta)
\eeq

To end this subsection, we would like to draw the attention that the analytical representations of both
 the (GK)  and (PK) coherent states are related through the Laplace transform $\cite{15}$

\section{Star product and Moyal Bracket}

\hspace{.3in}In this section, we introduce the star-product and Moyal bracket in coherent
states framework. Let us start by recalling the definition of star-product.
To every operator $A$  acting on the Hilbert space $ \mathcal{H}$ one can
associate a function ${\mathcal{A}} (z,\bar{z})$ on the complex plane as

\beq
{\mathcal{A}} (z,\bar{z})= <z|A|z>
\eeq
The associative star-product of two functions ${\mathcal{A}} (z,\bar{z})$ and
${\mathcal{B}} (z,\bar{z})$ is defined by $\cite{5}$

\beq
{\mathcal{A}} (z,\bar{z})*{\mathcal{B}} (z,\bar{z})=<z|AB|z>
\eeq
and then the corresponding Moyal bracket is given by

\beq
\{ {\mathcal{A}} (z,\bar{z}),{\mathcal{B}} (z,\bar{z}) \}_{M}=
{\mathcal{A}} (z,\bar{z})\star{\mathcal{B}} (z,\bar{z})-{\mathcal{B}}
 (z,\bar{z}) \star{\mathcal{A}} (z,\bar{z})= <z|[A,B]|z>.
\eeq
Using the identity resolution of the coherent states, the star-product equation (33) becomes

\beq
{\mathcal{A}} (z,\bar{z})\star{\mathcal{B}} (z,\bar{z})= \int d{\mu}
({\zeta}, \bar {\zeta}) <z|A|{\zeta}><{\zeta}|B|z>,
\eeq
which can be also written as
\beq
{\mathcal{A}} (z,\bar{z})\star{\mathcal{B}} (z,\bar{z})= \sum_{n,m}
 <z|\psi_n><\psi_n|AB|\psi_m><\psi_m|z>
\eeq
in terms of the function $<\psi_m|z>$ corresponding to the element $|\psi_m>$ of the Hilbert space
 ${\mathcal{H}}$. It follows that the Moyal bracket take the form
\beq
\{ {\mathcal{A}} (z,\bar{z}),{\mathcal{B}} (z,\bar{z}) \}_{M}= \sum_{n,m}
 <z|\psi_n><\psi_n|[A,B]|\psi_m><\psi_m|z>
\eeq

Analysing the relations (37), we see that there is a
correspondence between the structure relations of the operators
algebra of the and the $\star$ commutators, namely Moyal bracket,
of the elements generating the algebra of the functions on the
complex plane. This point will be examined through this
this section. \\
We note that the star-product (33) can be written in the integral  representation in terms of
the ordered exponential $\cite{5}$ :
\beq
\star= \int d{\mu}({\zeta}, \bar {\zeta}):\exp ( \frac{ \overrightarrow{ \partial}}{\partial{\eta}}(\zeta-\eta):
|<\eta|\zeta>|^{2} :\exp ((\bar{\zeta}-\bar{\eta})\frac{ \overleftarrow {\partial}}{\partial \bar{\eta}}:
\eeq
which not used in this work.

\subsection{Star product with Gazeau-Klauder Coherent states}

\hspace{.3in} In the GK Coherent states, the star-product take the
simple form

\beq {\mathcal{A}} (z,\bar{z})\star{\mathcal{B}}
(z,\bar{z})={\mathcal{N}} (|z|^2)^{-2}   \sum_{n,m}
{\mathcal{F}}_{n}(\bar{z}) <\psi_n|AB|\psi_m>{\mathcal{F}}_{m}(z)
\eeq Since the GK coherent states are  the eigenstates of the
annihilation operators $a^-$, there are a correspondence between
$a^-$ and the analytic function $z \to z$

\beq
<z|a^-|z>= z
\eeq
Then, the anti-analytic function  $z \to \bar {z}$  is the expectation  value of the
operators  $a^+$ over the coherent state $|z>$
\beq
 <z|a^+|z>=\bar{z}
\eeq
Furthermore, by using the definition of the star-product (39) one can obtain easily the
following relations

\beq 1\star1=1 \:\:\:\:\,\:\:\:\:\ 1\star z=z\star1=z
\:\:\:\:\,\:\:\:\:\ 1 \star \bar{z}=\bar{z} \star1=\bar{z} \eeq
and more generally, we have \beq z^{\star p}=z^{p}
\:\:\:\:\,\:\:\:\:\  \bar{z}^{ \star p}= \bar{z}^{p}
\:\:\:\:\,\:\:\ p \ge 0 \eeq where $ \theta^{\star p}=\theta \star
\theta \star \theta ...\star \theta$ ( p times, with $ \theta= z$
or $ \bar z)$. Based on the latter relations, one can evaluated
the star-product for any $z$-analytics and $\bar{z}$-antianalytics
functions which are given by \beq {\mathcal{A}}
(z)\star{\mathcal{B}} (z)= {\mathcal{A}} (z){\mathcal{B}} (z) \eeq
and \beq {\mathcal{A}} (\bar{z})\star{\mathcal{B}} (\bar{z})=
{\mathcal{A}} (\bar{z}){\mathcal{B}} (\bar{z}) \eeq We show also
that

\beq
\bar{z} \star z= <z|a^{+}a^-|z>= z \bar{z}=|z|^{2}
\eeq

Hence the star-product between two functions is reduced to the
ordinary one if the function in the right is analytic and the
function in the left is anti-analytic \beq {\mathcal{A}}
(\bar{z})\star{\mathcal{B}} (z)= {\mathcal{A}}
(\bar{z}){\mathcal{B}} (z) \eeq

For completeness, we will compute the star product of type $z \star \bar{z}$. From the previous
considerations it is easy to see that

\beq
z \star \bar{z}= <z|a^{-}a^{+}|z>= \bar{z} \star z -{\mathcal{G}} (z,\bar{z}).
\eeq
where

\beq
{\mathcal{G}} (z,\bar{z})=\frac{2a|z|^2}{r+1}  \frac{_{0}F_{1}(r+2,\frac{|z|^2}{a}) }{_{0}F_{1}(r+1,\frac{|z|^2}{a})}
+ (a+b)
\eeq
The function ${\mathcal{G}} (z,\bar{z})$ can be expressed as follows
\beq
{\mathcal{G}} (z,\bar{z})= 2az \frac{d}{dz} \sum_{n}{\mathcal{F}}_{n}(\bar z) {\mathcal{F}}_{n}(z) +a+b
\eeq
in terms of the functions $\{{\mathcal{F}}_{n} (z), n \in {\bf{N}} \}$.\\
One remark that the Moyal bracket  preserve the commutation relations of the algebra generated by
 $\{a^-, a^+ ,G \}$ as we have already mentioned . Indeed, in the operators language we have the following relations
\beq
[ a^-,a^+]=G(N) \:\:\:\:\:\ , \:\:\:\:\:\ [ a^{\pm},G(N)]={\pm}2aa^{\pm}
\eeq
which are  are expressed  in the language of Moyal bracket as

\beq
\{ z,\bar z \}_M={\mathcal{G}} (z,\bar{z}) \:\:\ , \:\:\  \{z,{\mathcal{G}} (z,\bar{z}) \}_M=2az \:\:\ , \:\:\
\{{\bar z},{\mathcal{G}} (z,\bar{z}) \}_M= -2a {\bar z}
\eeq
Using the previous results, one can show the following interesting relations
\beq
\bra{cccc}
\bar z \star{\mathcal{G}} (z,\bar{z})= {\mathcal{G}} (z,\bar{z}) + \bar{z}(a+b)-(a+b)$$ \\
 \\
{\mathcal{G}} (z,\bar{z}) \star \bar z= {\mathcal{G}} (z,\bar{z}) + \bar{z}(3a+b)-(a+b)$$ \\
\\
{\mathcal{G}} (z,\bar{z}) \star  z= {\mathcal{G}} (z,\bar{z}) + z(a+b)-(a+b)$$ \\
\\
z \star{\mathcal{G}} (z,\bar{z})= {\mathcal{G}} (z,\bar{z}) + z(3a+b)-(a+b)$$
\era
\eeq
which are useful to calculate in a complete way the star-product. Finally, using the relations (42), (43), (46), (52)
and (53), one can compute the star-product of any two functions ${\mathcal{A}} (z,\bar{z})$ and ${\mathcal{B}} (z,\bar{z}) $.
As illustration, let us give the following example : the star-product of  ${\mathcal{A}} (z,\bar{z})= \bar z $ and
${\mathcal{B}}(z,\bar{z})= \bar z z $ are given by

\beq
\bra{cc}
{\mathcal{A}} (z,\bar{z}) \star {\mathcal{B}} (z,\bar{z})= {\bar{z}}^2 z $$ \\
\\

{\mathcal{B}} (z,\bar{z}) \star {\mathcal{A}} (z,\bar{z})= {\bar{z}}^2 z +{\mathcal{G}} (z,\bar{z})+\bar z(a+b)-(a+b)
 , $$
\era
\eeq
and the corresponding Moyal bracket is

\beq
\{ {\mathcal{A}} (z,\bar{z}), {\mathcal{B}} (z,\bar{z}) \}_M=(a+b)-{\mathcal{G}} (z,\bar{z})-\bar z(a+b).
\eeq

In the particular case : $a=0$ and $b=1$ (i.e.,  the harmonic oscillator case), the function ${\mathcal{G}}(z,\bar{z})$ is equal
to unity and the relations (53) reduces to (42) ones. The relation (52) gives the well-known Moyal bracket constructed
using the coherent adapted to standard harmonic oscillator $\cite{5}$. \\

It is true that for the quantum systems considered in this work, we can define the Gazeau-Klauder coherent
states as well-as the Perelomov-Klauder ones. However, it should be noted that there exist some quantum systems
for which the Gazeau-Klauder coherent states can not be constructed due to the fact that the dimension of the Hilbert space
${\mathcal{H}}$ is finite like for instance a quantum system trapped in the Morse potential $\cite{16}$. In this
situation, the definition of the star-product discussed above can not be used. So, for this reason, we believe that
is interesting to introduce also the star-product in the Perelomov-Klauder coherent states.

\subsection{Star product with Perelomov-Klauder Coherent states}
\hspace{.3in}The Perelomov-Klauder Coherent states $|\zeta>$
equation (24) are not the eigenstates of the annihilation operator
$a^-$. In order to define the star-product and Moyal bracket, one
may ask if the analytic function $\zeta \to \zeta $ and the
anti-analytic function $ \zeta \to \bar {\zeta} $, can be defined
as the means values of some operators $A^-$ and $A^+$  acting on
the Hilbert space  ${\mathcal{H}}$ :

\beq
<\zeta|A^{-} |\zeta>= \zeta \:\:\:\:\  \:\:\:\:\ <\zeta|A^{+} |\zeta>= \bar{\zeta}
\eeq
Let us introduce the operators
\beq
A^{-} = a^- f(N)     \:\:\:\  \:\:\:\ A^{+} =f(N) a^+
\eeq
The operators $A^{-}$ and $A^{+}$ satisfy the relations (56), when  $f(N)$ is defined by

\beq
f(N)=\frac{N+1}{g(N+1)}
\eeq
where the function operator $g(N+l)$ acts in the Hilbert space as $|\psi_n>$

\beq
g(N+l)|\psi_n>=e_{n+l}|\psi_n>
\eeq
for $l \in {\bf N}$. The new operators $A^{-}$ and $A^{+}$ satisfy the following  relations
\beq
[A^{-}, A^{+}]= D(N)
\eeq
where the operator $D(N)$ is defined as a function of the operator $N$ by
\beq
D(N)= \frac {(N+1)^2}{g(N+1)} -\frac{N^2}{g(N)}
\eeq
One can show also that

\beq
A^-  D(N)= D(N+1)A^- \:\:\:\:\  \:\:\:\:\ A^+ D(N)=D(N-1)A^+
\eeq

For our purpose, we define the following functions
\beq
{\mathcal{D}}_{l}(\zeta,\bar{\zeta}) = <\zeta|D(N+l)|\zeta>
\eeq
which are useful, in the computation of the star-product based on the Perelomov-Klauder coherent states.
A straightfoward calculation leads to

\beq
{\mathcal{D}}_{l}(\zeta,\bar{\zeta})=(1-|\zeta|^{2})^{r+1} \sum_{n}
{\mathcal{G}}_{n}(\bar{\zeta}){\mathcal{G}}_{n}(\zeta)
\{ (n+l+2)^2 \frac{ e_{n+l+1}}{e_{n+l+2}^{2}}- (n+l+1)^2
\frac{e_{n+l}}{e_{n+l+1}^{2}}\}
\eeq
Using (64), one find the basic relations needed for a computation of star-product between any two functions ${\mathcal{A}}_{l}(\zeta,\bar{\zeta})$ and
${\mathcal{B}}_{l}(\zeta,\bar{\zeta})$. They are given

\beq
\bra{cccc}
1\star \zeta=\zeta \star 1=\zeta  \:\:\:\  \:\:\:\  1 \star \bar{\zeta}=\bar{\zeta} \star 1=\bar{\zeta}$$ \\
\\

\bar{\zeta} \star \zeta= \bar{\zeta} \zeta \:\:\:\:\:\:\:\:\:\  \:\:\:\:\:\:\:\:\:\ \zeta \star \bar{\zeta}= \bar{\zeta} \zeta+
{\mathcal{D}}_{0}(\zeta,\bar{\zeta}) $$ \\
\\

\zeta \star{\mathcal{D}}_{l}(\zeta,\bar{\zeta})= \zeta {\mathcal{D}}_{l+1}(\zeta,\bar{\zeta}) \:\:\:\:\:\:\  \:\:\:\:\:\:\
\bar {\zeta} \star{\mathcal{D}}_{l}(\zeta,\bar{\zeta})= \bar{\zeta} {\mathcal{D}}_{l}(\zeta,\bar{\zeta})$$ \\
\\

{\mathcal{D}}_{l}(\zeta,\bar{\zeta}) \star \zeta = \zeta {\mathcal{D}}_{l}(\zeta,\bar{\zeta})  \:\:\:\  \:\:\:\
{\mathcal{D}}_{l}(\zeta,\bar{\zeta}) \star \bar{\zeta} = \bar{ \zeta} {\mathcal{D}}_{l+1}(\zeta,\bar{\zeta})$$
\era
\eeq

As application,  we set ${\mathcal{A}} (\zeta,\bar{\zeta})= \bar {\zeta} $ and ${\mathcal{B}}
(\zeta,\bar{\zeta})= \bar{\zeta} \zeta $. The star-product is this case are given by

\beq
\bra{cc}
{\mathcal{A}}(\zeta,\bar{\zeta}) \star {\mathcal{B}} (\zeta,\bar{\zeta})= {\bar{\zeta}}^2 \zeta $$ \\
\\

{\mathcal{B}}(\zeta,\bar{\zeta}) \star {\mathcal{A}} (\zeta,\bar{\zeta})= {\bar{\zeta}}^2 \zeta
 + \bar{\zeta} {\mathcal{D}}_{0}(\zeta,\bar{\zeta})
 $$
\era
\eeq
where ${\mathcal{D}}_{0}(\zeta,\bar{\zeta})$ is defined by (64).\\
Contrary to the previous case ( one corresponding to GK coherent
states), the structure relations of the algebra $ \{ a^+, a^-, G
\}$ are not preserved by this star-product. However, one can see
that the Moyal bracket defined from Perelomov-Klauder
 coherent states preserve the commutation relations (60) and (62) of the algebra generated by
$\{ A^+, A^-, D(N) \}$. \\

For the harmonic oscillator case ($ a=0 , b=1$), we have $f(N)=1$, $D(N)=1$ and the operators $A^{\pm}$
reduce to the creation and annihilation operators of ordinary harmonic oscillator where the Moyal bracket is trivial.\\

We conclude that the star product in GK coherent states is in
general different from one obtained in the PK scheme, except for
the ordinary oscillator case.

\section {Concluding Remarks}

\hspace{.3in}In this work, we have introduced the star product and
the Moyal bracket in the coherent states framework corresponding
to exact solvable quantum systems, admitting a nonlinear spectra.
We have seen that in the PK coherent states case, the construction
becomes non-trivial, because the coherent states are not
eigenstates of the annihilation operator. This difficulty was
removing by introducing a new operator diagonalizing the PK
states. The fundamentals star-products, providing a complete way
to compute the Moyal bracket for any two functions, are given in
this work. The star product constructed for the standard harmonic
oscillator $\cite{5}$ was recorded as the particular case of our
approach. It is clear that there remain many problems for future
study. One of them would be the definition of star-product using
the coherent states for the Lie algebras and their supersymmetric
counterparts. Another would be a better understanding of the
relationship between the star product with GK coherent states and
one using the PK ones. We believe that such relation can be
established, because as we mentioned above, the analytical
representations of both coherent states are related through
Laplace transformation. Finally, it became apparent from this work
that the construction of the star product from the coherent states
involves a certain rule of correspondence between functions on
non-commuting operators and analytical functions; this
correspondence is similar to one between classical and quantum
mechanics. So, would be interesting, as suggested by one of the
referee of this paper, to show this calculus in use by studying an
exactly solvable quantum mechanics system cited in this work ( a
system trapped in P$\ddot{o}$schl-Teller potential, for instance).
 This matter in under consideration $\cite{17}$       \\
\\
${\bf{Acknowledgments}}$
\\

The authors would like to thank the Abdus Salam International
Center for theoretical Physics, Trieste, Italy,  for hospitality.
They also would like to thank the referees for their suggestions
and remakes.

\end{document}